\documentclass[11pt]{article}
\usepackage{multicol}
\usepackage{geometry}
\newgeometry{left=1.0in,right=1.0in,top=1.0in,bottom=1.0in}

\usepackage{float}
\floatstyle{boxed}
\restylefloat{figure}

\usepackage{mathpartir}
\usepackage{irule}
\usepackage{amssymb}
\usepackage{amsmath}
\usepackage{amsthm}
\usepackage{stmaryrd}
\usepackage{graphicx}
\usepackage{irule}

\usepackage{amssymb, amscd,txfonts,stmaryrd,setspace,mathtools,makecell}
\usepackage{enumerate}
\usepackage[usenames,dvipsnames]{xcolor}
\usepackage[bookmarks=true,colorlinks=true, linkcolor=MidnightBlue, citecolor=cyan]{hyperref}
\usepackage{lmodern}
\usepackage{graphicx,float}

\usepackage{titling}

\input xy
\xyoption{all} \xyoption{poly} \xyoption{knot}\xyoption{curve}
\usepackage{xypic,color}
\usepackage{tikz}

\newcommand{\comment}[1]{}

\def\tn{\textnormal}

\def\SEL*{\tn{SEL*}}

\def\hsp{\hspace{.3in}}

\def\LoopSchema{{\parbox{.5in}{\fbox{\xymatrix{\LMO{s}\ar@(l,u)[]^f}}}}}

\def\to{\rightarrow}

\def\|{{\;|\;}}
\def\m1{{-1}}

\newcommand{\LMO}[1]{\stackrel{#1}{\bullet}}

\newcommand{\LTO}[1]{\stackrel{\tt{#1}}{\bullet}}

\def\ullimit{\ar@{}[rd]|(.3)*+{\lrcorner}}
\def\urlimit{\ar@{}[ld]|(.3)*+{\llcorner}}
\def\lllimit{\ar@{}[ru]|(.3)*+{\urcorner}}
\def\lrlimit{\ar@{}[lu]|(.3)*+{\ulcorner}}
\def\ulhlimit{\ar@{}[rd]|(.3)*+{\diamond}}
\def\urhlimit{\ar@{}[ld]|(.3)*+{\diamond}}
\def\llhlimit{\ar@{}[ru]|(.3)*+{\diamond}}
\def\lrhlimit{\ar@{}[lu]|(.3)*+{\diamond}}
\newcommand{\clabel}[1]{\ar@{}[rd]|(.5)*+{#1}}
\newcommand{\TriRight}[7]{\xymatrix{#1\ar[dr]_{#2}\ar[rr]^{#3}&&#4\ar[dl]^{#5}\\&#6\ar@{}[u] |{\Longrightarrow}\ar@{}[u]|>>>>{#7}}}
\newcommand{\TriLeft}[7]{\xymatrix{#1\ar[dr]_{#2}\ar[rr]^{#3}&&#4\ar[dl]^{#5}\\&#6\ar@{}[u] |{\Longleftarrow}\ar@{}[u]|>>>>{#7}}}
\newcommand{\TriIso}[7]{\xymatrix{#1\ar[dr]_{#2}\ar[rr]^{#3}&&#4\ar[dl]^{#5}\\&#6\ar@{}[u] |{\Longleftrightarrow}\ar@{}[u]|>>>>{#7}}}

\newcommand{\arr}[1]{\ar@<.5ex>[#1]\ar@<-.5ex>[#1]}
\newcommand{\arrr}[1]{\ar@<.7ex>[#1]\ar@<0ex>[#1]\ar@<-.7ex>[#1]}
\newcommand{\arrrr}[1]{\ar@<.9ex>[#1]\ar@<.3ex>[#1]\ar@<-.3ex>[#1]\ar@<-.9ex>[#1]}
\newcommand{\arrrrr}[1]{\ar@<1ex>[#1]\ar@<.5ex>[#1]\ar[#1]\ar@<-.5ex>[#1]\ar@<-1ex>[#1]}

\newcommand{\Adjoint}[4]{\xymatrix@1{{#2}\ar@<.5ex>[r]^-{#1} &{#3} \ar@<.5ex>[l]^-{#4}}}

\newtheorem{theorem}{Theorem}[subsection]

\theoremstyle{remark}

\newtheorem{question}[theorem]{Question}
\newtheorem{guess}[theorem]{Guess}

\theoremstyle{definition}

\newcommand{\mainCatLarge}[1]{ 
	\stackrel{#1}{
		\parbox{4.1in}{\fbox{\parbox{4in}{\begin{center}\underline{{\tt Emp} manager worksIn $\simeq$ {\tt Emp} worksIn}\hsp  \underline{{\tt Dept} secretary worksIn $\simeq$ {\tt Dept}}\end{center}~\\\\\\
			\xymatrix@=10pt{&\LTO{Emp}\ar@<.5ex>[rrrrr]^{\tn{worksIn}}\ar@(l,u)[]+<8pt,10pt>^{\tn{manager}}\ar[dddl]_{\tn{first}}\ar[dddr]^{\tn{last}}&&&&&\LTO{Dept}\ar@<.5ex>[lllll]^{\tn{secretary}}\ar[ddd]^{\tn{name}}\\\\\\\LTO{FName}&&\LTO{LName}&~&~&~&\LTO{DName}
			}
		} }}
	}
}

\usepackage{diagrams}
\begin{document}

\title{QINL: Query-integrated Languages \\
{\large or, Co-(LINQ: Language-integrated Queries)}}
%\title{$\mcC_i$}
\author{Patrick Schultz, David I. Spivak
\\ Massachusetts Institute of Technology \\ {\sf \{pschultz, dspivak\}@math.mit.edu} 
\and
Ryan Wisnesky
\\
Categorical Informatics, Inc.
 \\ {\sf ryan@catinf.com}
}

\date{\today}

\maketitle
%\vspace*{-.5in}
%\begin{abstract}

%\end{abstract}
%\setlength{\columnsep}{.83cm}\begin{multicols}{2}

%\section{Introduction}
\vspace*{-.2in}
In this paper we describe an alternative solution to the impedance-mismatch problem between programming and query languages: rather than embed queries in a programming language, as done in LINQ~\cite{Tannen:1992:NEQ:645500.655920} systems, we embed programs  in a query language, and dub the result ``QINL''.  We have implemented our solution in a prototype software system, FQL, available at {\sf categoricaldata.net/fql.html}.  Because both LINQ and QINL extend a common language, type theory with products, we present this type theory first, then LINQ, and then QINL.

\section{Type theory with products}

To start our discussion, we now present a language, {\it type theory with products}, by the BNF grammar, typing rules, and equations in Figure 1.  In this language, $T$ represents types, $\Gamma$ represents typing contexts, $E$ represents expressions, and $x$ represents variables.  

\begin{figure}[h]
\caption{Type Theory with Products (TTP) over base types $\mathcal{T}$ and operations $\mathcal{E}$}
$$T ::=  1 \ | \ T \times T \ | \ \mathcal{T} \  \ \ \ \ \ \ \ \Gamma ::= - \ | \ \Gamma, v : T \ \ \ \ \ \ \
    E ::= x \ | \ () \ | \ (E, E) \ | \ E.1\  | \ E.2 \ | \ \mathcal{E}(E)$$
\setgroup{g1}
\irule{}{ }{\Gamma, x : T \vdash x : T}
\irule{}{\Gamma \vdash x : T \\ x' \neq x}{\Gamma, x' : T' \vdash x:T}
\irule{}{ }{\Gamma \vdash () : 1}
\irule{}{\Gamma \vdash E_1 : T_1 \\ \Gamma \vdash E_2 : T_2}{\Gamma \vdash (E_1, E_2) : T_1 \times T_2}
\irule{}{\Gamma \vdash E : T_1 \times T_2}{\Gamma \vdash E.1 : T_1}
\irule{}{\Gamma \vdash E : T_1 \times T_2}{\Gamma \vdash E.2 : T_2}
\irule{}{E_2 : T_1 \to T_2 \in \mathcal{E} \\ \Gamma \vdash E_1 : T_1 }{\Gamma \vdash E_2(E_1) : T_2}
\showrules{g1}

\setgroup{g2}
\irule{}{\Gamma \vdash E : 1}{\Gamma \vdash E = ()}
\irule{}{\Gamma \vdash E : T_1 \times T_2 }{\Gamma \vdash E = (E.1, E.2)}
\irule{}{\Gamma \vdash E_1 : T_1 \\ \Gamma \vdash E_2 : T_2}{\Gamma \vdash (E_1, E_2).1 = E_1}
\irule{}{\Gamma \vdash E_1 : T_1 \\ \Gamma \vdash E_2 : T_2}{\Gamma \vdash (E_1, E_2).2 = E_2}
%\irule{}{\Gamma \vdash E_1 = E_2 \in \mathcal{C}}{\Gamma \vdash E_1 = E_2}
\showrules{g2}

\end{figure}

A particular choice of $(\mathcal{T}, \mathcal{E})$ is called a {\it signature}, and the type-theory so generated is denoted TPP$(\mathcal{T}, \mathcal{E})$.  An {\it equational theory} over TPP$(\mathcal{T}, \mathcal{E})$ is a set of equations of the form $\Gamma \vdash E_1 = E_2$.

%\newpage
\section{LINQ}

LINQ~\cite{Tannen:1992:NEQ:645500.655920}, which stands for language-embedded query, is a technology that embeds comprehensions and other bulk-data operators into a general-purpose programming language.  LINQ technology is currently deployed in practical systems such as Microsoft's .NET, but the original LINQ system was conceived of at the University of Pennsylvania in the early 1990s and goes by the name ``nested relational calculus'' (NRC)~\cite{921235}.  Because of its simplicity, in this paper we will use the NRC as our canonical example of a LINQ system.  

The syntax and typing rules for the NRC are shown in Figure 2 and extend those of type theory with products from Figure 1.  In addition, the NRC posses an equational theory, not shown in Figure 2, that contains equations about sets, such as $E_1 \cup E_2 = E_2 \cup E_1$, for example.  For our purposes, it is enough to note that $\emptyset_T$ is intended to denote the empty set of type $T$, that $\cup$ denotes union, that $\{ E \}$ denotes the singleton set of only $E$, and that ${\sf for}$ denotes iteration.  Additionally, the NRC contains a boolean type {\sf Bool} and related operations.

\begin{figure}[h]
\caption{Nested Relational Calculus [extends Fig 1] (equations omitted)}
$$T ::= \ldots \ | \ {\sf Bool}  \ | \ {\sf Set} \ T  \ \ \ \ \ \ \ E ::= \ldots \ | \ \emptyset_T \ | \ E \cup E \ | \ \{ E \} \ | \ {\sf for} \ x \in E. \ E \ | \ {\sf if} \ E \ {\sf then} \ E \ {\sf else} \ E \ | \ {\sf t} \ | \ {\sf f} \ | \ E = E $$
\setgroup{g3}
\irule{}{ }{\Gamma \vdash \emptyset_T : {\sf Set} \ T}
\irule{}{\Gamma \vdash E_1 : {\sf Set} \ T \\ \Gamma \vdash E_2 : {\sf Set} \ T}{\Gamma \vdash E_1 \cup E_2 : {\sf Set} \ T}
\irule{}{\Gamma \vdash E : T}{\Gamma \vdash \{ E \} : {\sf Set} \ T}
\irule{}{\Gamma \vdash E_1 : {\sf Set} \ T_1 \\ \Gamma, x:T_1 \vdash E_2 : {\sf Set} \ T_2}
{\Gamma \vdash {\sf for} \ x \in E_1. E_2 : {\sf Set} \ T_2}
\irule{}{\Gamma \vdash E_1 : {\sf Bool} \\ \Gamma \vdash E_2 : T \\ \Gamma \vdash E_3 : T}
{\Gamma \vdash {\sf if} \ E_1 \ {\sf then} \ E_2 \ {\sf else} \ E_3 : T}
\irule{}{ }{\Gamma \vdash {\sf t} : {\sf Bool}}
\irule{}{ }{\Gamma \vdash {\sf f} : {\sf Bool}}
\irule{}{\Gamma \vdash E_1 : T \\ \Gamma \vdash E_2 : T}{\Gamma \vdash E_1 = E_2 : {\sf Bool}}
\showrules{g3}
\end{figure}

Many relational queries can be written in the NRC; for example, projecting the first column from a binary relation ($\pi_{col_1
}$), taking the cartesian product of two unary relations ($\times$), and selecting those rows from a binary relation that have the same values in both columns $(\sigma_{col_1=col_2})$:

\begin{figure}[h]
\caption{Some Relational Queries in NRC}
$$
\pi_{col_1} := I : {\sf Set} \ (T_1 \times T_2) \vdash {\sf for} \ x \in I. \{ x.1 \} : {\sf Set} \ T_1
$$
$$\times := I : ({\sf Set} \ T_1) \times ({\sf Set} \ T_2) \vdash {\sf for} \ x_1 \in I.1. \ {\sf for} \ x_2 \in I.2. \{ (x_1, x_2) \} : {\sf Set} \ (T_1 \times T_2)
$$
$$
\sigma_{col_1=col_2} :=  I : {\sf Set} \ (T_1 \times T_2) \vdash {\sf for} \ x \in I. {\sf if} \ x.1 = x.2 \ {\sf then} \ \{ x \} \ {\sf else} \ \emptyset_{T_1 \times T_2} : {\sf Set} \ (T_1 \times T_2)
 $$
\end{figure}

%In practice, a LINQ system extends the NRC with base types and operations. Indeed, a type of booleans and an if-then-else operation is required to implement relational selection, and many LINQ systems will also add first-class functions. For example, we might have
%\begin{figure}[h]
%\caption{An example LINQ extending the NRC (equations omitted)}
%$$T ::= \ldots \ | \ {\sf Bool} \ | \ {\sf Int} \ | \ {\sf String} \ | \ T \to T \ \ \ \ \ \ \ E ::= \ldots \ | \ {\sf bill} \ | \ {\sf length} \ | \ \lambda x:T. E \ | \ E(E) \ | \ {\sf if} \ E \ {\sf then} \ E \ {\sf else} \ E$$
%\setgroup{g4}
%\irule{}{ }{\Gamma \vdash {\sf bill} : {\sf String}}
%\irule{}{ }{\Gamma \vdash {\sf length} : {\sf String} \to {\sf Int}}
%\irule{}{\Gamma \vdash E_1 : T_1 \\ \Gamma \vdash E_2 : T_1 \to T_2}{\Gamma \vdash E_2(E_1) : T_2} 
%\irule{}{\Gamma, x:T_1 \vdash E:T_2}{\Gamma \vdash \lambda x:T_1. E : T_1 \to T_2}
%
%\showrules{g4}
%\end{figure}

To summarize, in the NRC, a database schema $S$ is a type, an instance on $S$ is a closed expression of type $S$, and a query $q$ from $S$ to another type $T$ is an open expression $I : S \vdash q : T$.  In this paper we will not dwell on the specifics of any particular LINQ system, but the following remarks about how particular LINQ systems are derived from the NRC are relevant:
\begin{itemize}
\item A particular LINQ system, such as .Net, can be modeled as the NRC generated by particular choice of $(\mathcal{T}, \mathcal{E})$, which we will denote as NRC$(\mathcal{T}, \mathcal{E})$.  A program (set of related definitions) in this LINQ system can be modeled as an equational theory over NRC$(\mathcal{T},\mathcal{E}$).   
\item Some expressive LINQ systems, such as Haskell, require not only a particular choice of $(\mathcal{T}, \mathcal{E})$, but a more expressive type theory as well - for example, a type theory containing exponential types (which model Haskell's first-class functions).  Other collection types besides sets, such as lists and bags, are also a common extension, as are recursive and polymorphic types.
\item The NRC, as defined above, cannot perform aggregation except through predefined operations (i.e., choosing $\mathcal{E}$ to contain e.g., ${\sf sum : Set \ Int \to Int}$).  However, the {\sf for} construct of the NRC can be extended to allow aggregation~\cite{755736}.  In this paper, we ignore aggregation.
\end{itemize}

%\newpage
\section{QINL}

Similarly to how we presented the NRC as the core of a LINQ system, we begin our discussion of QINL by presenting a language, FQL, that illustrates the core ideas of QINL.  For the purposes of this paper, we define FQL slightly differently than in the FQL IDE software tool.  Consider a particular signature $(\mathcal{T}, \mathcal{E})$.  An FQL {\it schema} is an equational theory over TTP$(\mathcal{T}, \mathcal{E})$. For example,

\begin{figure}[h]
\caption{Example FQL Schema [extends Figure 1]}
$$
\mathcal{T} := \{ {\sf String}, {\sf Int},  {\sf Emp}, {\sf Dept} \}
$$
$$\mathcal{E} := \{ {\sf length} : {\sf String} \to {\sf Int}, {\sf reverse} : {\sf String} \to {\sf String}, $$
$$
\ \ \ \ \ \ \ \ \ \ \ {\sf worksIn} : {\sf Emp} \to {\sf Dept}, {\sf manager} : {\sf Emp} \to {\sf Emp}, {\sf ename} : {\sf Emp} \to {\sf String} \}
$$
$$
 x : {\sf String} \vdash {\sf length}(x) = {\sf length}({\sf reverse}(x))
 \ \ \ \ \ \ \ \ x : {\sf String} \vdash x = {\sf reverse}({\sf reverse}(x))
 $$
 $$
  x : {\sf Emp} \vdash {\sf worksIn}(x) = {\sf worksIn}({\sf manager}(x)) 
$$
%T' := \{ {\sf Emp}, {\sf Dept} \} \ \ \ \ \ \ E' := \{ {\sf worksIn} : {\sf Emp} \to {\sf Dept}, {\sf manager} : {\sf Emp} \to {\sf Emp}, {\sf ename} : {\sf Emp} \to {\sf String} \} 
%$$ 
%$$
%C' := \{ x : {\sf Emp} \vdash {\sf worksIn}(x) = {\sf worksIn}({\sf manager}(x)) \}
%$$
\end{figure}

Because an FQL schema contains an arbitrary set of types, operations, and equations, each FQL schema is in fact a programming language -- indeed, FQL schemas can even be Turing-complete languages which have finite equational axiomatizations, such as the SK-combinatory algebra.  It is because programming languages live inside FQL schemas that we call FQL a QINL system, although perhaps SINL, or schema-embedded programming language, would be more accurate.  In contrast, in LINQ, schemas/types live inside an ambient programming language (the NRC).  

Suppose we wish to describe a database schema with entities $En$ and columns $Fk$ in an ambient programming language with types $Ty$ and operations $Fn$.  In the QINL approach, the schema would simply be the language TTP$(En \cup Ty, Fk \cup Fn)$.  In the LINQ approach, the schema would be a type, defined in terms of $En$ and $Fk$, in the language NRC$(Ty, Fn)$.  In other words, in QINL, an entity set $X$ is represented as a base type $X$; in LINQ, an entity set $X$ is represented as a expression of type ${\sf Set} \ Y$, for some appropriate $Y$.  QINL's  strategy of representing sets as types is common in type theory (e.g., Coq, Agda), but LINQ's strategy of representing sets as values/expressions is the dominant approach in database programming languages.  

Finally, we describe how instances an queries are handled in FQL. Let $(\mathcal{T}, \mathcal{E}, C)$ be an FQL schema.  Then an instance $I$ on this schema consists of, for each $T \in \mathcal{T}$, a set $I(T)$, and for each $E : T_1 \to T_2 \in \mathcal{E}$, a function $I(T) : I(T_1) \to I(T_2)$, such that all equations of $C$ are true in $I$.  Such instances can be described extensionally, for example as SQL database instances, or such instances can be described intensionally, for example, as initial models of set of equations; the FQL tool provides support for both kinds of descriptions.  For queries, FQL uses so-called data migration functors~\cite{Spivak:2012:FDM:2324905.2325108}: a {\it functor} $F$ between two FQL schemas $S$ and $S'$ is a function from base types in $S$ to base types in $S'$, and from base operations in $S$ to (open) expressions in $S'$, that preserves the equations of $S$.  Associated with $F$ are three operations, $\Delta_F, \Sigma_f, \Pi_F$, for migrating instances on $S$ to instances on $S'$, and vice versa; under suitable assumptions, compositions of these operations can be proved to be equivalence to unions of conjunctive relational queries~\cite{relfound}.  In fact, such queries may be written in a LINQ-like notation:

\begin{figure}[h]
\caption{Example FQL Query to find departments worked in by palindromic self-managers}
$$
{\sf for} \ e: {\sf Emp }\ {\sf where} \ {\sf manager}(e)=e \ {\sf and} \ {\sf reverse}({\sf ename}(e)) = {\sf ename}(e) \ {\sf return} \ {\sf worksIn}(e)
$$
\end{figure}
\newpage

\section{Conclusion}

To summarize, LINQ adds type constructors to product type theory so as to encode any schema as a type, and such that instances for a schema are terms of that type. In QINL, a schema becomes itself an extension of a product type theory, adding types for each entity, and an instance on a schema can be represented as a context $\Gamma$ in this extended type theory, together with equations in context $\Gamma$.

\bibliographystyle{plain}

\end{document}